\documentclass[11pt,twocolumn,twoside]{IEEEtran}
\usepackage{amsmath}
\usepackage[margin=0.75in,headheight=0.45in]{geometry}
\usepackage{epsfig}
\usepackage{amsfonts}
\usepackage{amssymb}
\usepackage{fancyhdr}
\usepackage{tikz}
\include{graphicsx}
\usepackage{graphicx}
\usepackage{epstopdf}
\usetikzlibrary{shapes,decorations,arrows,calc,arrows.meta,fit,positioning}
\tikzset{
    -Latex,auto,node distance =1 cm and 1 cm,semithick,
    state/.style ={ellipse, draw, minimum width = 0.7 cm},
    point/.style = {circle, draw, inner sep=0.04cm,fill,node contents={}},
    anticausal/.style={dashed},
    el/.style = {inner sep=2pt, align=left, sloped}
}

\begin{document}
\title{\vspace{0.2in}\sc A direct approach to detection and attribution of climate change}
\author{Enik\H{o} Sz\'ekely$^{1, *}$, Sebastian Sippel$^{2}$, Reto Knutti$^{2}$, Guillaume Obozinski$^{1}$, Nicolai Meinshausen$^{2}$\thanks{$^1$Swiss Data Science Center, ETH Z\"urich and EPFL, Switzerland $^2$ETH Z\"urich, Switzerland, Corresponding author: E. Sz\'ekely, eniko.szekely@epfl.ch }}

\maketitle
\thispagestyle{fancy}
\begin{abstract}
We present here a novel statistical learning approach for detection and attribution (D\&A) of climate change. Traditional optimal D\&A studies try to directly model the observations from model simulations, but practically this is challenging due to high-dimensionality. Dimension reduction techniques reduce the dimensionality, typically using empirical orthogonal functions, but as these techniques are unsupervised, the reduced space considered is somewhat arbitrary. Here, we propose a supervised approach where we predict a given external forcing, e.g., anthropogenic forcing, directly from the spatial pattern of climate variables, and use the predicted forcing as a test statistic for D\&A. We want the prediction to work well even under changes in the distribution of other external forcings, e.g., solar or volcanic forcings, and therefore formulate the optimization problem from a distributional robustness perspective.

\end{abstract}

\section{Introduction}
Traditional Detection and Attribution (D\&A) methods quantify the connection between observations and model simulated responses to different external forcings \cite{AllenTett1999, HegerlZwiers2011, BindoffEtAl2013}. While detection aims to find if there is a change in the observations that cannot be explained by internal variability alone, attribution tries to assign the detected change to a particular external forcing or a combination of forcings. D\&A studies first reduce the dimensionality by projecting onto the space spanned by the first few empirical orthogonal functions (EOFs)/principal components (PCs), and then regression is used in this reduced space to estimate the scaling factors \cite{AllenTett1999}. One of the issues with the dimension reduction is that the procedure is unsupervised, and the resulting reduced space depends on the precise form of the dimension reduction technique. EOFs/PCs reduce the dimension by finding the few leading eigenvectors/fingerprints that maximize the variance, but the choice of the number of eigenvectors remains subjective. 

We propose here a proof of concept in a perfect model scenario where we predict directly the radiative forcing, e.g., anthropogenic forcing, in a supervised way, and use the predicted radiative forcing as a test statistic for D\&A. The supervised setting allows us to find the projection of interest that best explains the radiative forcing, and avoids the arbitrariness of unsupervised dimension reduction as preprocessing step. To ensure that the results are robust to changes in the distribution of other external forcings, e.g., solar or volcanic forcing, we formulate the optimization problem from a distributional robustness perspective. We aim to find a robust estimator for a whole class/set of distributions, not only for the target population distribution. The set of distributions will be given by climate interventions in model simulations, e.g., control runs, Representative Concentration Pathways (RCPs), and the class of shift interventions on the external forcings, e.g., natural, solar, or volcanic forcing. This work fits into the emerging framework of data-driven approaches for detection and attribution using data assimilation \cite{HannartEtAl2016} or statistical and machine learning \cite{BarnesEtAl2018, SippelEtAl2019_dailydetection}.

\section{Methodological framework}

\subsection{Traditional Detection and Attribution}

Traditional D\&A studies first extract the fingerprint of external forcings from model simulations driven with the respective forcing by averaging across a large number of runs to reduce the influence of internal variability \cite{HegerlEtAl1996, SanterEtAl2018}. In addition, the so-called optimal D\&A studies project both the model simulations and the observations onto the space spanned by the first few EOFs of a set of (unforced) control simulations that feature only internal climate variability \cite{AllenTett1999}. 

Let $X_{obs}^{EOF} \in \mathbb{R}^{T \times d}$ and $X_M^{EOF} \in \mathbb{R}^{T \times d}$ be the projection of the observations $X_{obs}$ (e.g., temperature, precipitation) and climate responses from model simulations $X_M$ (e.g., temperature, precipitation) onto the first $d$ EOFs of internal (natural) variability, where $T$ is the number of simulated years. $X_M$ consists of $k$ sets of forced simulations, typically $k=2$ with $X_M = \{X_{ANT}, X_{NAT}\}$, i.e., simulations with only anthropogenic and only natural forcing, respectively. The scaling parameters $\alpha \in \mathbb{R}^k$ corresponding to the set of forced simulations, e.g., $\alpha_{ANT}$ and $\alpha_{NAT}$, are estimated from regression in the space of the EOFs:
\begin{equation}
\label{eq:traditionalDA}
\hat{X}_{obs}^{EOF} = \sum_{i=1}^k \alpha_i X_M^{i, EOF}.
\end{equation}
\noindent The magnitude and confidence intervals of the scaling factors $\alpha_i$ indicate how much of the signal in the observations can be attributed to each particular forcing.

\subsection{Data-driven Detection and Attribution}

Let $Y \in \mathbb{R}^{n}$ be the simulated radiative forcing (e.g., anthropogenic, volcanic, solar, GHG, CO2), and $X \in \mathbb{R}^{n \times p}$ the matrix of specific climate variable measurements (e.g., temperature, precipitation, humidity) from climate model simulations in any given year, where $n$ is the number of samples (i.e., the total number of simulated years across all simulations) and $p$ the dimensionality of the data (the number of features or spatial grid cells). The radiative forcing is the net change in the energy balance of the Earth system due to some imposed perturbation \cite{MyhreEtAl2013}. Here, the matrix $X$ is obtained by concatenating the different model simulation runs, and not through averaging as in the case of $X_M$ from \eqref{eq:traditionalDA}, therefore $n=m \times T$, where $m$ is the number of model simulation runs and $T$ the number of years simulated for each model.

The alternative data-driven D\&A approach that we propose here predicts the external forcing $y \in Y$ directly from model simulations $x \in X$. Let $f_{\beta}(x)$ be the function that predicts $y$ and is parameterized by $\beta$. The parameters $\beta$ are estimated by minimizing a loss function $l ( y ,  f_{\beta}(x) ) $ over the population drawn from some target population distribution $(x,y) \sim P$:

\begin{equation}
\label{eq:populationLoss}
\hat{\beta} = \operatorname*{argmin}_{\beta} \mathbb{E}_{(x,y) \sim P} [l ( y ,  f_{\beta}(x) ) ].
\end{equation}

The model used $f_{\beta}(\cdot)$ can be a linear or nonlinear (kernel) regression model, a random forest or a deep neural network \cite{BarnesEtAl2018}. The estimator in \eqref{eq:populationLoss} only optimizes over one target population distribution $(x, y)\ {\sim}\ P$, however as we will see in the next sections, changes in the distribution of the data, e.g., stronger solar or volcanic forcing, can lead to poor prediction results. Our goal here is to protect ourselves against such distributional changes in the external forcings and optimize over a whole class of distributions in order to ensure robustness (for details see Sects.\ \ref{sect:distrRobust} and \ref{sect:anchorRegr}). 

In the statistical model from \eqref{eq:populationLoss}, the parameters $\beta$ are learned from climate model simulations, and can be used to predict the external forcing from observations: 
\begin{equation*}
\hat{y}_{obs} = f_{\beta}(x_{obs}),
\end{equation*}
\noindent where $x_{obs}$ are the full observational maps, and $\hat{y}_{obs}$ is the predicted observed forcing. We focus in this short paper on a perfect model scenario where we predict data from model simulations and leave the prediction of the observations for future work.

Traditional D\&A tries to explain an \textit{observed }climate pattern as a function of \textit{modelled} climate patterns from simulations driven with different external forcings as in eq. \eqref{eq:traditionalDA}. However, due to the high-dimensionality, this step cannot be performed directly on the original data. D\&A therefore first extracts the fingerprints using an EOF analysis and the regression is performed in the EOF space. Our direct approach is an alternative to this step of fingerprint extraction. Instead of extracting the fingerprint using an (unsupervised, and therefore somewhat arbitrary) EOF analysis, we extract the fingerprint using directly the information contained in the radiative forcing in a supervised way. We note that our goal is not to predict the radiative forcing, but to use it to extract the fingerprint and, as explained in the following, to define a test statistic used for detection and attribution.

In the data-driven supervised approach that we propose, \textit{detection} is done by testing against the null hypothesis that the predicted forcing does not differ from internal (natural) variability, i.e., the predicted forcing is not significantly different from zero. Practically, this is done by considering the predicted forcing $\hat{y}$ (either from model simulations in a perfect model scenario, or from observations) as a one-dimensional vector test statistic and computing the confidence intervals of the prediction. Detection occurs if the test statistic $\hat{y}$ is outside the pre-industrial range, i.e., the confidence intervals do not contain zero; and \textit{attribution} is established if the true forcing lies within the confidence intervals of the predicted forcing.

\subsection{Distributional robustness}
\label{sect:distrRobust}

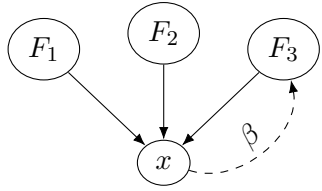
\begin{figure} [ht]
\begin{center}
\begin{tikzpicture}
    \node[state] (x) at (0,0) {$x$};
    \node[state] (y1) [above left =of x] {$F_1$};
    \node[state] (y2) [above =of x] {$F_2$};
    \node[state] (y3) [above right =of x] {$F_3$};

    \path (y1) edge (x);
    \path (y2) edge (x);
    \path (y3) edge (x);

    \path[anticausal] (x) edge[bend right =60] node[el, above] {$\beta$} (y3);
\end{tikzpicture}
\caption{Causal diagram for the effect (plain arrows) of external forcings $F_1, F_2,$ and $F_3$, e.g., solar, volcanic and anthropogenic forcing, on the climate response $x$, e.g., temperature, and the regression problem (dashed arrow) for predicting $y = F_3$.}
\label{fig:causal}
\end{center}
\end{figure}

The climate response $x$, e.g., temperature, is potentially influenced by multiple external forcings. Let us consider that we have three external forcings, e.g., solar, volcanic and anthropogenic forcings (see causal diagram in Fig.\ \ref{fig:causal}), and let's say we want to predict the anthropogenic forcing $y = F_3$. The diagram can be extended to include other forcings if necessary, or to split the anthropogenic forcing into its constituent parts, e.g., GHG, CO$_2$.

As discussed in the previous section, if we were to predict the anthropogenic forcing using the regression model in \eqref{eq:populationLoss}, we would only optimize over $\beta$ for the observed distribution $(x, y)\ {\sim}\ P$. But, instead of seeking an estimator $f_{\beta}(\cdot)$ which is just a good predictor of the value of the forcing for the given distribution, we actually would like $f_{\beta}(\cdot)$ to capture the specific effect of the targeted forcing regardless of the strength of the other forcings. In other words, we would like to guarantee good prediction results even under distributional changes and we want the null distribution of the test statistic (in case of detection) to be valid even under changed solar/volcanic forcing.

The class of distributions $\mathcal{Q}$ over which we want to achieve robustness is generated both by interventions on the climate models (e.g., control runs, RCPs, anthropogenic runs, natural runs), and the class of shift interventions, i.e., interventions that shift the value of a variable in a given direction  \cite{Meinshausen2018}, on the external forcings. For example, in the graph from Fig.\ \ref{fig:causal}, the shift distributions $Q \in \mathcal{Q}$ are obtained by shifting the forcing $F_1$ or $F_2$, e.g., solar or volcanic forcing. 

The distributionally robust form of the estimator in \eqref{eq:populationLoss} is given by 
\begin{equation}
\label{eq:distrRobust}
\hat{\beta} = \operatorname*{argmin}_{\beta} \sup_{Q \in \mathcal{Q}} \mathbb{E}_{(x,y) \sim Q} [l ( y ,  f_{\beta}(x) ) ],
\end{equation}
that optimizes over a whole class of distributions $\mathcal{Q}$ instead of just a single target population distribution $P$ \cite{Meinshausen2018, Buehlmann2018}. Distributional robustness is formulated here as a worst-case scenario, where solving for the most difficult case guarantees good prediction results for unseen future distributions. 

\subsection{Anchor regression}
\label{sect:anchorRegr}

In the example from Fig.\ \ref{fig:causal}, we would like to protect ourselves against changes in the distribution of the solar forcing $F_1$, the volcanic forcing $F_2$, or both. We call these variables \textit{anchors}, and in a linear setting where $f_{\beta}(x) = x^T\beta$ and the loss function is the least squares empirical risk $\sum_{i=1}^n l ( y_i ,  f_{\beta}(x_i) ) = \Vert Y - X \beta \Vert_2^2$, we use anchor regression \cite{RothenhaeuslerEtAl2019} to achieve the distributional robustness from \eqref{eq:distrRobust}. 

Let the anchor variables be $ A \in \mathbb{R}^{n \times q}$, where $n$ is the number of samples and $q$ is the number of anchors. The robust estimator of anchor regression is given by
\begin{equation}
\label{eq:anchor}
\hat{\beta}^{\gamma} = \operatorname*{argmin}_{\beta} \Vert (I_n - \Pi_A)(Y - X \beta) \Vert_2^2 + 
					\gamma \Vert \Pi_A(Y-X \beta) \Vert_2^2,		
\end{equation}

\noindent where $\Pi_A \in \mathbb{R}^{n \times n}$ is the matrix that projects on the column space of $A$, i.e., $\Pi_A = A(A^T A)^{-1} A^T$, $I_n \in \mathbb{R}^{n \times n}$ is the identity matrix, and $\gamma$ is the ``causal'' regularization parameter that gives the strength of the shift intervention on the anchor variable. The causal regularization encourages orthogonality (or uncorrelatedness) of the residuals with the anchor variable. For the graph in Fig.\ \ref{fig:causal}, this ensures that the prediction accuracy remains good even if the strength of the solar or volcanic forcing changes. For $\gamma = 1$, the projection of the residuals on $\Pi_A$ vanishes between the two terms and anchor regression coincides with ordinary least squares:
\begin{equation}
\label{eq:dataDrivenDA}
\hat{\beta}^{1} = \operatorname*{argmin}_{\beta} \Vert Y - X \beta \Vert_2^2,
\end{equation}
where the parameters $\beta \in \mathbb{R}^{p}$ are the maps of regression coefficients (that can be interpreted in a more traditional sense in climate science as ``fingerprints''). As the solution of both ordinary least squares from \eqref{eq:dataDrivenDA} and anchor regression from \eqref{eq:anchor} can be prone to overfitting, we include a regularization term in the optimization problem. Because we want to ensure the smoothness of the maps $\beta$, we will use ridge (Tikhonov) regularization \cite{HastieEtAl2009}, and the estimator for the model in \eqref{eq:anchor} can be written as ridge regression on a transformed data set:
\begin{equation}
\label{eq:ridge}
\hat{\beta}^{\gamma} = \operatorname*{argmin}_{\beta} \Vert \tilde{Y} - \tilde{X} \beta \Vert_2^2 + \lambda \Vert \beta \Vert_2^2,
\end{equation}


\noindent where $\lambda$ is the regularization parameter that controls the bias-variance tradeoff, and $\tilde{X} = (I_n-\Pi_A)X+\sqrt{\gamma} \Pi_AX$ and $\tilde{Y} = (I_n-\Pi_A)Y+\sqrt{\gamma} \Pi_AY$ are the transformed data sets. The second term in \eqref{eq:ridge} penalizes large regression coefficients, and handles the multicollinearity of the predictors. 

Anchor regression finds the direction $\beta$ that explains the component of the climate response to the forcing of interest, e.g., the anthropogenic forcing, that is orthogonal to other components that are (possibly) common in the response to other forcings. This allows us to do attribution of the detected change in the climate variable: if the projection on $\beta$ (predicted forcing) is similar enough to the forcing of interest (true forcing), the change can be attributed to the respective forcing. 

\begin{figure*}[ht!]
\begin{center}
		\includegraphics[trim = {3cm 2.5cm 3cm 2cm}, width = 0.95\textwidth]{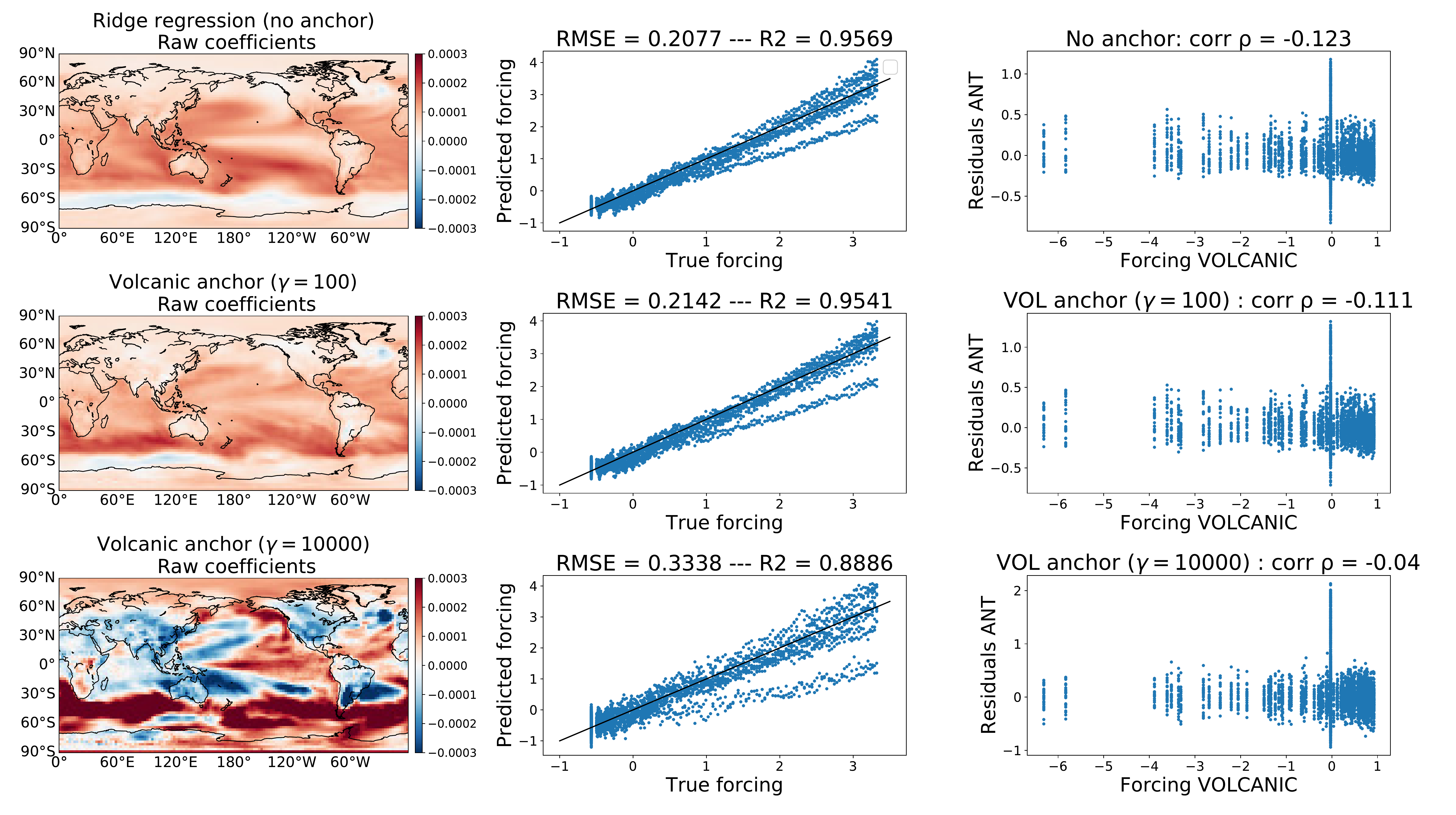}
	\end{center}	\caption{Prediction of the anthropogenic forcing using anchor regression with the volcanic anchor. The first row shows the results for ridge regression, while the second and third row show results for anchor regression with two values of the causal regularization parameter $\gamma$.}
\label{fig:predictionTOTAL}
\end{figure*}


\section{Data}

We use data from climate model simulations from CMIP5 (Climate Model Intercomparison Project) \cite{TaylorEtAl2012} and consists of control runs and Representative Concentration Pathways (RCPs) \cite{MossEtAl2010} -- RCP 2.6, RCP 4.5, RCP 6, RCP 8.5 -- that outline plausible forcing trajectories throughout the 21st century that are used to drive climate model simulations. Here we use 42 control run simulations and 40 RCP 8.5 model simulations, so 82 model simulations from 21 climate models. Each model simulation has an annual resolution and runs for 231 years from 1870 to 2100. In total there are $n = 82 \times 231 = 18,942$ samples. The samples are two-dimensional spatial maps, and the spatial resolution is $p = 144 \times 72 = 10,368$ dimensions.

We first subtract the mean of the period 1870-1920 from each model individually in order to remove model biases in mean temperature, and then standardize the data prior to regression analysis. The regularization coefficient $\lambda$ is chosen by cross validation with the folds built model-wise, i.e., we make sure that data from the same climate model falls in the same fold. We use here $k = 3$ folds. Likewise, the splitting into training and testing is also done model-wise. This ensures that we are testing only on full models that have not been seen during training. The data is split into 75\% of models for training, and the remaining 25\% of models for testing.

\section{Experiments and results}

We report results for the prediction of the anthropogenic forcing using the volcanic forcing as anchor (Fig.\ \ref{fig:predictionTOTAL}). The first row shows the results of ridge regression (anchor regression with $\gamma=1$), while the second and third row show the results for anchor regression with two different values of the ``causal'' parameter $\gamma$. The first column shows the raw coefficients $\beta$ of the regression; the second column shows the prediction results together with the RMSE and R2 score for each case; and the last column plots the residuals against the anchor variable. We would like to obtain residuals that are uncorrelated with (or ideally independent from) the anchor to guarantee good prediction results even if the anchor changes. Constraining with the volcanic anchor slightly lowers the prediction accuracy (lower R2 and higher RMSE), however it also protects against a strong volcanic forcing. The correlation of the residuals with the anchor (last column) goes to zero as we increase the parameter $\gamma$ from anchor regression. In the middle column we observe one testing model that behaves fairly different from the rest of the models (represented by the points that deviate the most from the black line). The raw coefficients with low causal regularization have mostly positive values indicating that all grid points contribute to explain the warming, but rely more on the tropical oceans because they have less variability than polar regions. Also land areas are chosen less than adjacent ocean regions, and the ENSO region is not chosen because it shows variability that is irrelevant w.r.t. the anthropogenic forcing. With the increase in the causal regularization parameter, the contrast in the maps also increases, as the coefficients give more weight to regions that play a role in explaining the anthropogenic forcing, but not the volcanic forcing. We note here that this approach is not intended to find estimates of the historical radiative forcing. Instead, we find through regression a spatial pattern that captures the (linear) relationship between the temperature and the existing radiative forcing estimates, and subsequently (in future work) we will use these spatial patterns to predict the observed radiative forcing for detection and attribution.

\begin{figure}[ht!]
\begin{center}
		\includegraphics[trim = {1cm 2cm 1cm 0cm}, width = 0.4\textwidth]{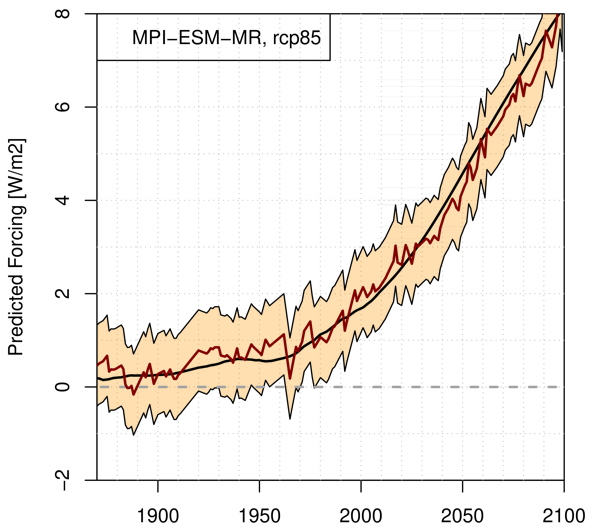}
	\end{center}	\caption{Detection and attribution in a supervised setting (see text for details).}
\label{fig:confInterval}
\end{figure}

Fig.\ \ref{fig:confInterval} shows how detection and attribution work using an RCP run with the anthropogenic forcing as target variable. The signal is detected starting around 1990, i.e., the confidence intervals after this time don't contain zero anymore, and we can attribute the signal to the anthropogenic forcing because the true forcing (black) lies within the confidence intervals ($\pm 2 \sigma$) of the predicted forcing (red). We compute the confidence intervals for each model separately by scaling the standard deviation of the residuals of the prediction for each value of the forcing by the standard deviation of the residuals for the corresponding scenario, i.e., RCP vs control runs. The confidence intervals are defined here with respect to the residuals of the prediction, and therefore they allow us to use them for hypothesis testing in detection and attribution. However we note that this definition of confidence intervals should not be confused with the standard definition in the statistical learning literature where the confidence intervals are defined with respect to the mean value of the predictions.

\section{Conclusion}

We have introduced a novel supervised statistical learning approach for studying the detection and attribution of climate change that protects against distributional changes in the external forcings. The class of distributions that we would like to protect ourselves against is generated by both interventions on the climate models and implicit shift interventions via the anchor method on the external forcings. In future work we plan to extend the framework to other forcings and go towards independence of the residuals with the anchor instead of just orthogonality. Another future direction is to incorporate temporal information into our framework, using for example Takens embedding (time-delay coordinates) \cite{Takens1981, SauerEtAl1991}. As the climate response to external forcings is not instantaneous, such information might help disentangle the different forcings which act on different timescales.

\section*{Acknowledgments}
This work was partly funded by the Swiss Data Science Center within the project ``Data-science informed attribution of changes in the hydrological cycle'' (DASH, C17-01). We thank Urs Beyerle for the preparation and maintenance of the CMIP5 data. We thank two anonymous reviewers for their valuable comments. 

\bibliographystyle{ieeetr}
\bibliography{bibliography}

\end{document}